\documentclass{appolb}
\usepackage{graphicx}

\newcommand{\Bb}{{\bar{B}}}

\newcommand{\Bsb}{{\bar{B}_s}}

\newcommand{\BSb}{{\bar{B}^*}}

\newcommand{\BsSb}{{\bar{B}_s^*}}

\newcommand{\SU}{{\rm SU}}

\newcommand{\MeV}{{\,\rm MeV}}

\newcommand{\ignore}[1]{} 

\begin{document}
\title{Odd-parity dynamically generated baryon resonances with beauty flavor%
\thanks{Presented at the Workshop "Excited QCD 2013", Bjelasnica Mountain, Sarajevo, Bosnia and Herzegovina, February 3-9, 2013}%
\
\thanks{
Supported by Spanish Ministerio de Econom{\'\i}a y Competitividad
(FIS2011-28853-C02-02, FIS2011-24149, FPA2010-16963),
Junta de Andalucia (FQM-225), Generalitat Valenciana (PROMETEO/2009/0090) and EU HadronPhysics2 project (grant 227431). O. R. acknowledges support from the Rosalind Franklin Fellowship.
L. T. acknowledges support from RyC Program, and FP7-PEOPLE-2011-CIG (PCIG09-GA-2011-291679).}
}
\author{O.~Romanets
\address{KVI,
  University~of~Groningen, Zernikelaan~25,~9747AA~Groningen, The~Netherlands}
\\
C.~Garc{\'\i}a-Recio,  L.~L.~Salcedo
\address{Departamento~de~F{\'\i}sica~At\'omica, Molecular~y~Nuclear, and Instituto
  Carlos I de F{\'\i}sica Te\'orica y Computacional, Universidad~de~Granada,
  E-18071~Granada, Spain}
\\
J.~Nieves
\address{Instituto~de~F{\'\i}sica~Corpuscular~(centro~mixto~CSIC-UV),
  Institutos~de~Investigaci\'on~de~Paterna, Aptdo.~22085,~46071,~Valencia,
  Spain}
\\
L.~Tolos
\address{ Institut~de~Ci\`encies~de~l'Espai~(IEEC/CSIC),
  Campus~Universitat~Aut\`onoma~de~Barcelona, Facultat~de~Ci\`encies,
  Torre~C5,~E-08193~Bellaterra,~Spain, \\
Frankfurt Institute for Advanced Studies, \\
Johann Wolfgang Goethe University, Ruth-Moufang-Str.~1,
60438 Frankfurt am Main, Germany}
}
\maketitle
\begin{abstract}
We study baryon resonances with heavy flavor in a molecular approach, thus as dynamically generated by baryon-meson scattering. 
This is accomplished by using a unitary coupled-channel model taking, as bare interaction, the extension 
of the Weinberg-Tomozawa $\SU (3)$ Lagrangian.
A special attention is payed to the inclusion of heavy-quark spin symmetry and to 
the study of the generated baryon resonances that complete the heavy-quark spin multiplets.
Our model reproduces the $\Lambda_b(5912)$ and $\Lambda_b(5920)$ baryons, which were recently observed by the LHCb collaboration. 
According to our analysis, these two states are heavy-quark spin symmetric partners. We also make predictions for few $\Xi_b$ 
baryon resonances, which belong to the same SU(3)$\times$HQSS multiplets as the $\Lambda_b$ particles.

\end{abstract}
  
\section{Introduction}
Using $pp$ collision data 
the LHCb Collaboration~\cite{Aaij:2012da} has
reported the discovery of two narrow states, observed in the
$\Lambda^0_b\pi^+\pi^-$ spectrum, with masses $5911.97 \pm
0.12\,(stat)\,\pm 0.02\,(syst)\,\pm 0.66\,(\Lambda^0_b \, {\rm mass})\MeV,$
and $5919.77\pm 0.08\,(stat)\,\pm 0.02\,(syst)\, \pm
0.66\,(\Lambda^0_b\, {\rm mass})\MeV$. 
These states are interpreted as
the orbitally-excited $\Lambda^0_b(5912)$ and $\Lambda^0_b(5920)$
bottom baryon resonances, with spin--parity $J^P=1/2^-$ and
$J^P=3/2^-$, respectively.   The limits on the natural widths 
of these states are
$ \Gamma_{\Lambda^0_b(5912)} \leq 0.83\MeV$ and $
\Gamma_{\Lambda^0_b(5920)} \leq 0.75\MeV$ at the 95\% confidence level ~\cite{Aaij:2012da}.
These results  reported by the LHCb collaboration 
are in good agreement with an old prediction by Capstick and Isgur~\cite{Capstick:1986bm}.
Their relativistic quark model predicts $5912 \MeV$ and $5920 \MeV$ for the
masses of the lightest orbitally-excited
states. However, the same model yields a mass
of the ground state $\Lambda^0_b$ ($J^P=1/2^+$) which is about $35 \MeV$ smaller
than the measured value~\cite{Beringer:1900zz}. 
At the same time some other existing quark-models predictions  
\cite{Garcilazo:2007eh,Ebert:2007nw,Karliner:2008sv,Roberts:2007ni} differ by few
tenths of MeV from the LHCb experimental ones.
\ignore{
More recently, Garcilazo et
al.~\cite{Garcilazo:2007eh} have also presented  
results from a constituent quark model scheme. 
They adjusted the mass of the $\Lambda^0_b$ ground state and
predicted the masses of the $J^P=1/2^-$ and $3/2^-$ orbitally excited
$\Lambda_b$ states, which turned out to be around $30$ $\MeV$ lower
than the LHCb experimental values. Note however that the masses predicted
in ~\cite{Garcilazo:2007eh} are in turn 20-$30\MeV$ higher than those obtained
in other schemes based also on the relativistic quark model
\cite{Ebert:2007nw}, or on the color hyperfine interaction
\cite{Karliner:2008sv} or on the heavy quark effective theory
\cite{Roberts:2007ni}.
 More recently, in  \cite{Ortega:2012cx} heavy baryonic
  resonances $\Lambda_b$ ($\Lambda_c$) with $J^P=3/2^-$ are studied in
  a constituent quark model as a molecular state composed by nucleons
  and $\bar{B^*}$ ($D^*$) mesons.
}

We adopt a different approach and describe these 
states
as dynamically generated resonances obtained within a unitarized baryon-meson
coupled-channel scheme.
\ignore{It is known that some baryon states can be
constructed as a $qqq$ state in a quark model, and simultaneously as a dynamically
generated resonance in a baryon-meson coupled-channel description (that is a
$qqq -q \bar q $ molecular state) \cite{Klempt:2009pi}. Though, some of their
properties might differ. It is thus interesting to consider 
both points of view in order to get in the future a
joint or integral description of hadronic resonances in terms of quarks and
hadrons degrees of freedom.}
The unitarization in coupled-channels has proven to be very
successful in describing some of the existing experimental data. 
Thus, there have been different successful studies 
based on the chiral perturbation theory amplitudes for scattering
of $0^-$ octet Goldstone bosons off baryons of the  $1/2^+$ nucleon octet in the charmless sector
(e.g.~\cite{Kaiser:1995eg,Oset:1997it,
Tolos:2000fj,GarciaRecio:2003ks}).
Unitarized coupled-channel methods have been further extended to the baryon-meson
sector with charm degrees of freedom
in several ways.
In some works, 
e.g.~\cite{Tolos:2007vh,JimenezTejero:2009vq} a bare  baryon-meson interaction is used, 
saturated with the $t$-channel exchange of vector mesons between pseudoscalar mesons
and baryons.
Some other works \cite{Haidenbauer:2007jq,Haidenbauer:2010ch} are based on the
J\"ulich meson-exchange model and those in \cite{Wu:2010jy,Wu:2010vk,Oset:2012ap}
make use of the
hidden gauge formalism, and the same approach has been extended to the bottom sector in \cite{Wu:2010rv}. 
An extended Weinberg-Tomozawa (WT) interaction implementing heavy-quark spin symmetry (HQSS) is applied in
\cite{GarciaRecio:2008dp,Romanets:2012hm,Gamermann:2010zz,GarciaRecio:2013hc} for the charm sector, 
while in \cite{GarciaRecio:2013} it was also used for 
bottom-flavored baryon states.
In this talk, we review some of the results of \cite{GarciaRecio:2013}.
%
\ignore{
Of special importance are the symmetries that are implemented in the quark or
hadronic models. Typically, while hadronic models pay an special attention
to chiral symmetry, quark models usually implement heavy-quark spin symmetry
(HQSS).
HQSS is a direct consequence of QCD ~\cite{Isgur:1989vq,Neubert:1993mb,MW00}. It states that the
interaction dependent on the spin state of the heavy quark is of
$O(\Lambda_{{\rm QCD}}/m_Q)$, and so suppressed in the infinite quark mass limit. For
instance, the vector and pseudoscalar mesons with a bottom quark, which only differ
in how the spins of light and heavy quarks are coupled, form a doublet of HQSS
and would be degenerate in the infinite mass limit. So, HQSS requires the
pseudoscalar $\Bb~(\Bsb)$ meson and the $\BSb~(\BsSb)$ meson, its vector
partner, to be treated on an equal footing.  On the other hand, chiral
symmetry fixes the lowest order interaction between Goldstone bosons and other
hadrons in a model independent way; this is the Weinberg-Tomozawa (WT)
interaction.  Thus, it is appealing to have a predictive model for four
flavors including all basic hadrons (pseudoscalar and vector mesons, and
$\frac{1}{2}^+$ and $\frac{3}{2}^+$ baryons) which reduces to the WT
interaction in the sector where Goldstone bosons are involved and which
incorporates HQSS in the sector where bottom quarks participate.
}

We use hadronic degrees of freedom in a unitarized baryon-meson
coupled-channel calculation. We rely on a tree-level interaction
that embodies the approximate pattern of chiral symmetry, when Goldstone
bosons are involved, and HQSS
when heavy hadrons are present. Moreover, it
enjoys spin-flavor symmetry in the light 
 ($u,~d,~s$) flavor  sector \cite{GarciaRecio:2013}.
\ignore{
The scheme has been
successfully used for describing odd parity $s$-wave light
flavor~\cite{Gamermann:2011mq} and
charm~\cite{GarciaRecio:2008dp,Gamermann:2010zz,Romanets:2012hm} baryon resonances.  Indeed,
the model naturally explains the overall features (masses, widths and main
couplings) of the corresponding resonances ($\Lambda_c^+(2595)$
and $\Lambda_c^+(2625)$) that appear in the charm sector
($C=1$)~\cite{GarciaRecio:2008dp,Romanets:2012hm}. Further predictions of this
model for the $C=-1$ sector can be found in \cite{Gamermann:2010zz,GarciaRecio:2011xt}.
}

\section{Theoretical framework}
We follow the approach previously applied for charm systems in 
Refs.~\cite{GarciaRecio:2008dp,Romanets:2012hm,Gamermann:2010zz}. 
We consider baryon resonances with one bottom quark ($B=-1$),
in particular $\Lambda_b$ states with strangeness, isospin and spin-parity 
$(S,I,J^P)=(0,0,1/2^-)$ quantum numbers, $\Lambda_b^*$ with $(0,0,3/2^-)$,
$\Xi_b$ with $(-1,1/2,1/2^-)$, and $\Xi_b^*$ with $(-1,1/2,3/2^-)$.

Baryon-meson pairs with the same $SIJ$ quantum numbers span
the coupled-channel space. 
We study $s$-wave tree-level amplitudes between
two coupled channels $ij$, which are given by:
\begin{equation}
V_{ij}^{SIJ} =
D_{ij}^{SIJ}\,\frac{2\sqrt{s}-M_i-M_j}{4f_if_j} 
\sqrt{\frac{E_i+M_i}{2M_i}}\sqrt{\frac{E_j+M_j}{2M_j}}, 
\label{eq:pot}
\end{equation}
where $\sqrt{s}$ is the center of mass (C.M.) energy of the system; $E_i$ and
$M_i$ are, respectively, the C.M. energy and mass of the baryon in the channel
$i$; and $f_i$ is the decay constant of the meson in the $i$-channel.  The
masses of the baryons and
the mesons and the meson decay
constants used in this work can be found in~\cite{Romanets:2012hm, GarciaRecio:2013}. 
The coefficients $D_{ij}^{SIJ}$ come from the underlying spin-flavor
extended WT structure of the couplings of our model
\cite{GarciaRecio:2008dp}. The various exact
symmetries referred above (chiral, spin-flavor and HQSS) apply only to the
coefficients $D_{ij}^{SIJ}$, while physical masses and decay meson constants
are used throughout when solving the coupled-channel equations. 
\ignore{Tables for the
coefficients can be found in the Appendix B of
Ref.~\cite{Romanets:2012hm}. The coefficients to be used for the $B=-1$ sector
(one bottom quark interacting with light quarks) are identical to those for
$C=1$ (one charm quark interacting with light quarks) with obvious renaming of
the heavy hadrons.  The universality of the interactions of heavy quarks,
regardless of their concrete (large) mass, flavor and spin state, follows from QCD
~\cite{Isgur:1989vq,Neubert:1993mb,MW00} and it is automatically implemented in our model. Let
us remark that such emerging heavy spin-flavor symmetry, which becomes exact
in the infinitely heavy quark limit, is different from the approximate SU(6)
or light spin-flavor symmetry, also implemented in our model.}
Further, we 
solve the Bethe-Salpeter equation in the complex plane,
which provides the $T$-matrix as
\begin{equation}
T^{SIJ}=(1-V^{SIJ}G^{SIJ})^{-1}V^{SIJ}\label{eq:bse},
\end{equation}
where $G^{SIJ}$ is a diagonal matrix containing the two particle propagator
for each channel. The two particle propagator diverges
logarithmically, thus the loop is renormalized by a
subtraction  constant (see \cite{Romanets:2012hm} for 
discussion about the use of this method) such that
%
$G_{ii}^{SIJ}=0, ~{\rm at}~ \sqrt{s}=\mu^{SI}$. 
%
To fix the subtraction point $\mu^{SI}$ we consider
all sectors with a common $SI$ and different $J$ and all the corresponding
channels. Then $\mu^{SI}$ is taken as
$\sqrt{m_{\rm{th}}^2+M_{\rm{th}}^2}$, where $m_{\rm{th}}$ and $M_{\rm{th}}$
are, respectively, the masses of the meson and the baryon producing the lowest
threshold for a given $SI$ sector.

The dynamically-generated baryon resonances are obtained as poles of the
scattering amplitudes in each of the  $SIJ$ sectors.  
The poles of the
scattering amplitude on the first Riemann sheet that appear on the real
axis below the threshold are interpreted as bound states, and those found
on the second Riemann sheet below the real axis and above threshold are
identified with resonances. 
%
The position $\sqrt{s_R}$ of the pole
on the complex energy plane
\ignore{Close to the pole, the $T$-matrix behaves
as 
\begin{equation} \label{Tfit} T^{SIJ}_{ij} (s) \approx \frac{g_i
e^{i\phi_i}\,g_je^{i\phi_j}}{\sqrt{s}-\sqrt{s_R}} \,.  \end{equation} %
}
$\sqrt{s_R}=M_R - \rm{i}\, \Gamma_R/2$ provides the mass ($M_R$) and the width
($\Gamma_R$) of the resonance, 
and the couplings to the different baryon-meson channels 
are obtained from the residues of the scattering amplitude
around the pole.


\section{Dynamically generated $\Lambda_b$ and $\Xi_b$ resonances}
\subsection{$\Lambda_b$ and $\Lambda_b^*$ states}

Our model generates four $\Lambda_b$ resonances.
Three lowest lying $\Lambda_b$
states have masses of $5880$ and $5949\MeV$ ($J^P=1/2^-$) and $5963\MeV$
($J^P=3/2^-$).  As one can expect, the situation in the $J=1/2^-$ channel
keeps a close parallelism with that of the $\Lambda_c(2595)$ resonance in the
charm sector~\cite{GarciaRecio:2008dp, Romanets:2012hm}. For both heavy
flavors the structure obtained mimics the well-known two-pole pattern of the
$\Lambda(1405)$,
e.g. \cite{GarciaRecio:2003ks}. Thus, we find that
the state at $5880\MeV$ strongly couples to the $N \bar B$ and $N \bar B^*$
channels, with a negligible $\Sigma_b \pi $ coupling, while the $5949$
MeV state has
a sizable coupling to this latter channel. On the other hand, the $J^P=3/2^-$
state at $5963\MeV$ is generated mainly by the ($N\bar{B}^*$, $\Sigma_b^* \pi$)
coupled-channel dynamics. This state is the bottom counterpart of the
$\Lambda(1520)$ and $\Lambda_c^*(2625)$ resonances.

In order to achieve a better agreement with
the $\Lambda_b(5912)$ and $\Lambda_b(5920)$ states reported by the
LHCb Collaboration, we have slightly changed
the value of the subtraction point in this sector, namely we
have set the baryon-meson loop to be zero at the C.M. energy
$\sqrt{s}=\mu$ given by
$\mu^2 =\alpha~(M_{\Sigma_b}^2+m_\pi^2) $. 
For $\alpha=0.967$ we find two poles above the $\Lambda^0_b\pi\pi$ threshold,
with masses $5910.1\MeV$ ($J^P=1/2^-$) and $5921.5\MeV$ ($J^P=3/2^-$), which
admit a natural identification with the two $\Lambda_b$ resonances
 observed in~\cite{Aaij:2012da}. 
Their masses lie below all two-body channels thresholds
considered in our calculations.
The lowest in mass $J^P=1/2^-$ $\Lambda_b$ resonance is located at 5797.6~MeV, 
and decays radiatively to $\Lambda_b \gamma$.
The other spin-1/2 baryon resonance in this sector has a mass of $6009.3 \MeV$,
which is above the $\Sigma_b \pi$ threshold, but with less than 0.5~MeV width
due to small coupling.
%

We find that the states $\Lambda_b(5912)$ and
$\Lambda^*_b(5920)$ are 
HQSS
partners.
These two
states would be part of a {\bf $3^*$} irreducible representation (irrep) of
SU(3), embedded in a {\bf 15} irrep of SU(6)
(this was checked using the adiabatic symmetry breaking described in~\cite{Romanets:2012hm}). 
Thus, the light quark
structure of these two states is the same,
and the coupling of the $b$-quark spin with the spin
of the light degrees of freedom yields $J=1/2$ and $J=3/2$, and 
$\Lambda_b(5912)$ and $\Lambda^*_b(5920)$ states form an approximate
degenerate doublet.

\ignore{  ###
Comparison of Table \ref{tablambdab} with the Table III of
Ref.~\cite{Romanets:2012hm} in the charm sector, shows that states with the
same group labels in both tables are the heavy flavor counterpart of each
other. In particular, the $\Lambda_b(5920)$ resonance is the bottom version of
$\Lambda_c(2625)$ one, while the $\Lambda_b(5912)$ 
would not be the counterpart of the $\Lambda_c(2595)$ resonance,  
but it would be of the second charmed state that appears around 2595
MeV, and that gives rise to the two pole structure~\cite{Romanets:2012hm} mentioned above. The same conclusion follows from inspection of their
couplings: the $\Lambda_c(2595)$ couples weakly to $\Sigma_c \pi$ while the
coupling to $\Sigma_b\pi$ is sizable for the $\Lambda_b(5912)$ state.
}

As it was already mentioned in the introduction, 
different quark models
\cite{Capstick:1986bm,Garcilazo:2007eh,Ebert:2007nw,Karliner:2008sv,Roberts:2007ni}
have also conjectured the existence of one or more excited $\Lambda_b(1/2^-)$
and $\Lambda_b(3/2^-)$ states. The early
work of Capstick and Isgur \cite{Capstick:1986bm} generated the first two
excited $\Lambda_b(1/2^-)$ and $\Lambda_b(3/2^-)$ states with masses that are
particularly
in very good agreement with the ones observed by the LHCb
collaboration, but the ground state $\Lambda_b(1/2^+)$ mass in this
scheme is below the experimental one. Our model reproduces the experimental
$\Lambda_b(5912)$ and $\Lambda_b(5920)$ with an alternative explanation of
their nature as molecular states.

\subsection{$\Xi_b$ and $\Xi_b^*$ states}

In this work we found
three $\Xi_b$ and one $\Xi^*_b$, which belong to the
same SU(3)$\times$HQSS  multiplets of the $\Lambda_b$ and $\Lambda^*_b$
states introduced in the previous subsection.
For the subtraction point in this sector we use $\mu^2 = M_{\Xi_b}^2+m_\pi^2$.
We find a $\Xi_b(5874)$ spin-1/2 state, which does not have possible hadronic decays,
and thus its main decay mode is $\Xi_b \gamma$. 
Further we find $\Xi_b(6035.4)$ $J^P=1/2^-$ and  $\Xi_b(6043.3)$ $J^P=3/2^-$,
both with negligible width,
which lie above the $\Xi_b \pi$ threshold. 
The fourth resonance we find in this sector is $\Xi_b(6072.8)$ with a small width of 
0.3~MeV and two possible strong decay modes $\Xi_b \pi$ and $\Xi_b' \pi$.
We find
that $\Lambda_b(5797.6)$ and $\Xi_b(5874)$ belong to the same irreducible
representation, and similarly the $\Lambda_b(6009.3)$ and $\Xi_b(6072.8)$
states. Also, the pair $\Xi_b(6035.4)$ and $\Xi_b^*(6043.3)$, in the {\bf 15}
irrep of SU(6), form the HQSS doublet related by SU(3) to the doublet formed by
the $\Lambda_b(5910.1)$ and $\Lambda_b^*(5921.5)$ states.

\ignore{ ###
The three $\Xi_b$ and one $\Xi^*_b$ states have also partners in the
charm sector. We find that states with the same group labels are the heavy
flavor counterpart of each other, as already noted for the $\Lambda_b$ and
$\Lambda^*_b$ sectors. By comparing Table~\ref{tabxib} with Table V of
Ref.~\cite{Romanets:2012hm}, we see that the HQSS partners in the charm
sector coming from the ${\bf 15}$ representation, $\Xi_c(2772.9)$ and
$\Xi^*_c(2819.7)$, are the bottom counterparts of the $\Xi_b(6035.4)$ and
$\Xi^*_b(6043.3)$ states. Moreover, the charmed $\Xi_c(2699.4)$ and $\Xi_c(2775.4)$
resonances are analogous to the $\Xi_b(5874)$ and
$\Xi_b(6072.8)$ ones in the bottom sector,
respectively.
}
 None of these beauty-flavored states have been seen 
experimentally yet. Schemes based on quark models
\cite{Capstick:1986bm,Garcilazo:2007eh,Ebert:2007nw,Karliner:2008sv,Roberts:2007ni}
predict $\Xi_b(1/2^-)$ and $\Xi_b(3/2^-)$ states with similar masses 
to our estimates, though there exist some differences between the
various predictions.   

\ignore{ ###
Fig.~\ref{fig:fig1} shows a summary of the masses of the predicted
$\Lambda_b(1/2^-)$, $\Lambda_b(3/2^-)$, $\Xi_b(1/2^-)$ and
$\Xi_b(3/2^-)$ states with respect to the mass of the ground state
$\Lambda_b$, together with several thresholds for possible two- and
three-body decay channels. The experimental $\Lambda_b^0(5912)$ and
$\Lambda_b^0(5920)$ of LHCb are given for reference. 
}

%

\section{Summary}
We have analyzed odd-parity baryons with one bottom quark by means of
a unitarized
baryon-meson coupled-channel model which implements heavy-quark spin symmetry.
In our model
pseudoscalar and vector heavy mesons are treated on an equal
footing. We rely on a relatively simple tree-level interaction already used
in the charm sector \cite{GarciaRecio:2008dp,Romanets:2012hm}. 
This interaction has the virtue of embodying the
approximate patterns of chiral symmetry, when Goldstone bosons are involved, and
HQSS when heavy hadrons are present.
 The experimental states $\Lambda^0_b(5912)$ and
$\Lambda^0_b(5920)$ reported by the LHCb collaboration are obtained as
dynamically generated baryon-meson molecular states.  Within our
scheme, these states are identified as HQSS partners, which naturally
explains their approximate mass degeneracy. Other $\Lambda_b$
states coming from the same attractive SU(6) $\times$ HQSS
representations are also analyzed. 
Mass and decay mode predictions are also obtained for $\Xi_b(1/2^-)$ and
$\Xi_b(3/2^-)$ resonances, which belong to the same SU(3)
multiplets as the $\Lambda_b(1/2^-)$ and $\Lambda_b(3/2^-)$ states,
and the related underlying symmetry structure is analyzed.

\ignore{ ???
\section{Acknowledgments}
We thank E. Ruiz Arriola for discussions and Anton Poluektov for useful
information on the LHCb experimental setup. This research was supported by DGI
and FEDER funds, under contracts FIS2011-28853-C02-02, FIS2011-24149,
FPA2010-16963 and the Spanish Consolider-Ingenio 2010 Programme CPAN
(CSD2007-00042), by Junta de Andaluc{\'\i}a grant FQM-225, by Generalitat
Valenciana under contract PROMETEO/2009/0090 and by the EU HadronPhysics2
project, grant agreement n. 227431. O.R. wishes to acknowledge support from
the Rosalind Franklin Fellowship. L.T. acknowledges support from Ramon y Cajal
Research Programme, and from FP7-PEOPLE-2011-CIG under contract
PCIG09-GA-2011-291679.
}


\end{document}